\documentclass[preprint,aps,showpacs,nofootinbib,preprintnumbers,amsmath,amssymb]{revtex4-1}
\usepackage{epsfig}
\usepackage{subfigure}
\usepackage{dcolumn}
\usepackage{bm}
\usepackage[usenames ,dvipsnames]{xcolor}
\usepackage{slashed}
\usepackage{graphicx,color}

\begin{document}
\title{Simultaneous extractions of $|V_{ub}|$ and $|V_{cb}|$\\ 
with only the exclusive $\Lambda_b$ decays}

\author{Y.K. Hsiao and C.Q. Geng}
\affiliation{
School of Physics and Information Engineering, Shanxi Normal University, Linfen 041004, China\\
Department of Physics, National Tsing Hua University, Hsinchu, Taiwan 300}
\date{\today}

\begin{abstract}
We perform the  simultaneous $|V_{ub}|$ and $|V_{cb}|$ extractions
with only the exclusive $\Lambda_b$ decays of $\Lambda_b\to (p,\Lambda_c^+)\mu\bar \nu_\mu$,
$\Lambda_b\to p\pi^-$ and $\Lambda_b\to \Lambda_c^+ (\pi^-,  D^-)$.
We obtain that $|V_{ub}|=(3.7\pm 0.3)\times 10^{-3}$ and $|V_{cb}|=(45.9\pm 2.7)\times 10^{-3}$.
Our value of $|V_{ub}|$  is larger than that of $(3.27\pm 0.15\pm 0.16\pm 0.06)\times 10^{-3}$, 
previously extracted by the LHC Collaboration from the exclusive $\Lambda_b$ decays also, but 
nearly identical to  $(3.72\pm 0.19)\times 10^{-3}$ from the exclusive $B$ decays.
On the other hand, our extracted result of $|V_{cb}|$ favors
the value of $(42.2\pm 0.8)\times 10^{-3}$ from the inclusive $B$ decays.

\end{abstract}

\maketitle
\section{introduction}
The Cabibbo-Kobayashi-Maskawa (CKM) quark mixing matrix elements,
$V_{ub}$ and $V_{cb}$, in the standard model have been studied extensively in the literature~\cite{pdg}.
It is known that there are long-standing discrepancies for their determinations 
from the exclusive and inclusive $B$ decays.
Explicitly, it is given that~\cite{pdg}
\begin{eqnarray}\label{data1}
|V_{ub}|_{in}&=&(4.49\pm 0.16^{+0.16}_{-0.18})\times 10^{-3}\;\,,\nonumber\\
|V_{ub}|_{ex}&=&(3.72\pm 0.19)\times 10^{-3}\;\,,\nonumber\\
|V_{cb}|_{in}&=&(42.2\pm 0.8)\times 10^{-3}\;\,,\nonumber\\
|V_{cb}|_{ex}&=&(39.2\pm 0.7)\times 10^{-3}\;\,,
\end{eqnarray}
where
 the subscripts ``$in$'' and  ``$ex$'' stand for the extractions from
 the inclusive and exclusive $B$ decays, respectively.  
Clearly, it is important to have some examinations besides  the $B$ meson ones, such as those from the 
inclusive and exclusive $\Lambda_b$ decays.
Indeed, with the branching ratios of
the $\Lambda_b\to \Lambda_c^+\ell\bar \nu_\ell$ and 
$\Lambda_b\to\Lambda_c^+ M_{(c)}$ decays, 
where $M=(\pi^-,K^-)$ and $M_c=(D^-,D^-_s)$,
$|V_{cb}|$ is extracted to be $(44.6\pm 3.2)\times 10^{-3}$~\cite{Hsiao:2017umx}.
The extraction is
in accordance with the recent studies, where $|V_{cb}|_{ex}$
has been raised to agree with $|V_{ub}|_{ex}$~\cite{Bigi:2017njr}.
In addition, an extraction of $|V_{ub}|/|V_{cb}|$ from $\Lambda_b\to p\mu\bar \nu_\mu$ and 
$\Lambda_b\to \Lambda_c^+ \mu\bar \nu_\mu$  has been performed by the LHCb Collaboration~\cite{Aaij:2015bfa}.

In Ref.~\cite{Aaij:2015bfa}, although 
the absolute branching ratio of  $\Lambda_b\to p\ell\bar \nu_\ell$ has not been observed,
the ratio of 
the partial branching fractions of $\Lambda_b\to p\mu\bar \nu_\mu$ and 
$\Lambda_b\to \Lambda_c^+ \mu\bar \nu_\mu$ 
has been used, which is given by~\cite{Aaij:2015bfa}
\begin{eqnarray}\label{data2}
{\cal R}_{ub/cb}\equiv
\frac
{{\cal B}(\Lambda_b\to p \mu\bar \nu_\mu)_{q^2>15\,\text{GeV}^2}}
{{\cal B}(\Lambda_b\to \Lambda_c^+\mu\bar \nu_\mu)_{q^2>7\,\text{GeV}^2}}
=(1.00\pm 0.09)\times 10^{-2}\,,
\end{eqnarray}
with the selected transferred energy squared $q^2$ regions 
of $q^2>15$~GeV$^2$ and 7~GeV$^2$.
According to the theoretical calculation,
formulated by
\begin{eqnarray}\label{Rub}
\frac{
{\cal B}_{th}(\Lambda_b\to p \mu \bar \nu)_{q^2>15\,{\text{GeV}^2}}}
{{\cal B}_{th}(\Lambda_b\to\Lambda_c^+\mu\bar \nu_\mu)_{q^2>7\,{\text{GeV}^2}}}
=\frac{|V_{ub}|^2/|V_{cb}|^2}{R_{FF}}\,,
\end{eqnarray}
with $R_{FF}=0.68\pm 0.07$ 
as the ratio of the $\Lambda_b\to \Lambda_c$ and $\Lambda_b\to p$ transition form factors,
calculated by the lattice QCD (LQCD)~\cite{Detmold:2015aaa},
it is presented that $|V_{ub}|/|V_{cb}|=0.083\pm 0.004\pm 0.004$. 
The simplest way to extract $|V_{ub}|$ is by
putting the existing value of $|V_{cb}|$ into $|V_{ub}|/|V_{cb}|$. 
With $|V_{cb}|=(39.5\pm 0.8)\times 10^{-3}$ 
from the $B\to D^{(*)}\ell\bar \nu_\ell$ decays~\footnote{The data from PDG of the 2014 edition~\cite{2014PDG}.}
it was extracted by LHCb that
$|V_{ub}|=(3.27\pm 0.15\pm 0.16\pm 0.06)\times 10^{-3}$~\cite{Aaij:2015bfa},
which is even 2$\sigma$ lower than $|V_{ub}|_{ex}$ in Eq.~(\ref{data1}),
indicating that the discrepancy 
between the exclusive and inclusive $|V_{ub}|$ determinations
cannot be alleviated in the exclusive $\Lambda_b$ decays.
On the other hand, by using
$|V_{cb}|=(44.6\pm 3.2)\times 10^{-3}$  from
the exclusive $\Lambda_b$ decays~\cite{Hsiao:2017umx}
into the most recent value of $|V_{ub}|/|V_{cb}|=0.095\pm 0.005$ in the PDG~\cite{pdg}
that combines the values from both the exclusive $B$ and $\Lambda_b$ decays,
it gives rise to  $|V_{ub}|=(4.3\pm 0.4)\times 10^{-3}$,
and draws a different conclusion.

The two $|V_{ub}|$ values deviate with each other. Besides,
none of them can be claimed to be purely extracted 
from the exclusive $\Lambda_b$ decays.
In this study, we propose to have a complete global fit
with the currently existing data in the exclusive $\Lambda_b$ decays,
such as ${\cal R}_{ub/cb}$ and $\Lambda_b\to p \pi^-$,
performing the simultaneous $|V_{ub}|$ and $|V_{cb}|$  determinations.
 Since $R_{FF}$ will be no longer taken as 
an independent theoretical input in the extraction,
the uncertainties due to the theoretical calculations of 
the $\Lambda_b\to (\Lambda_c^+,p)$ form factors
can be reduced.
In addition, the possible data correlations should be carefully considered. 
We will also take into account 
the recently updated ${\cal B}(\Lambda_c^+\to p K^-\pi^+)$ data.
As a result, we can unambiguously extract $|V_{ub}|$ and $|V_{cb}|$
by fitting with only the exclusive $\Lambda_b$ decays,
which are connected by the two ratios in Eqs.~(\ref{data2}) and (\ref{Rub}),
to be regarded as the independent examination other than the B meson ones.

The paper is organized as follows. In Sec. II, we show the formalism. 
We give our numerical results and discussions in Sec. III.
We conclude in Sec. VI.

\section{Formalism}
\begin{figure}[t!]
\centering
\includegraphics[width=2.1in]{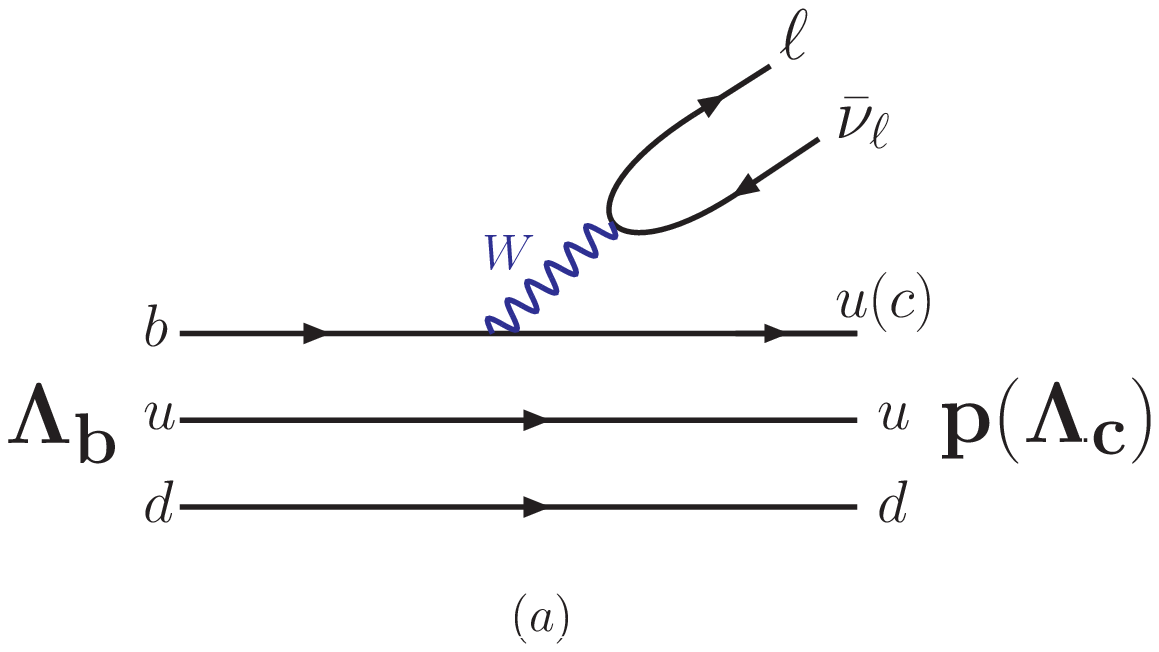}
\includegraphics[width=2.1in]{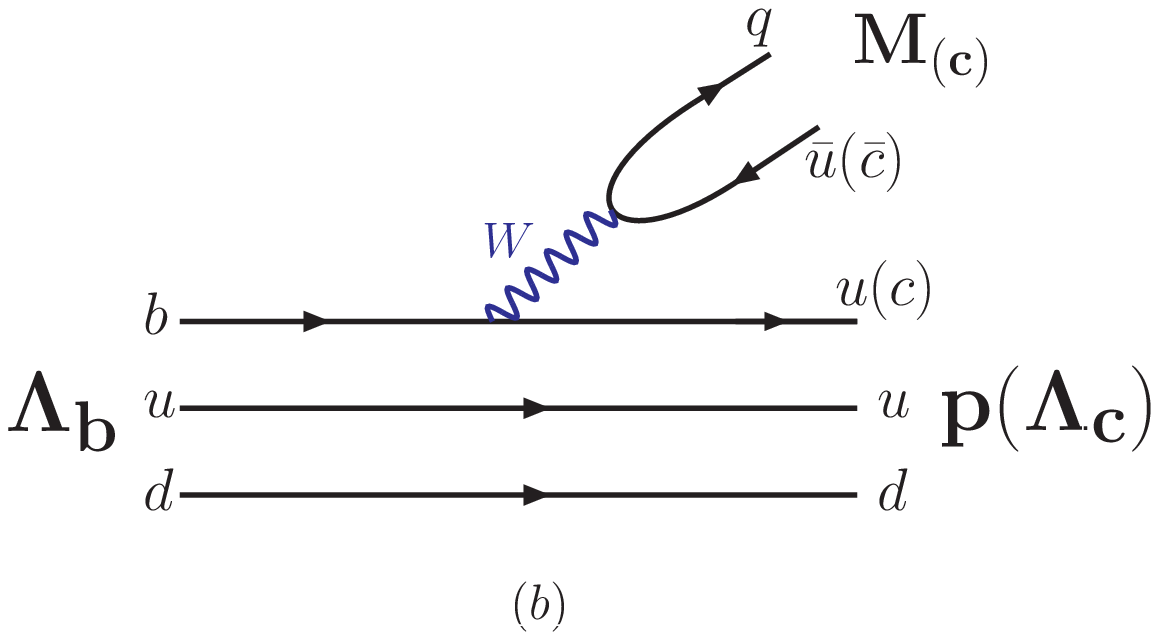}
\includegraphics[width=2.1 in]{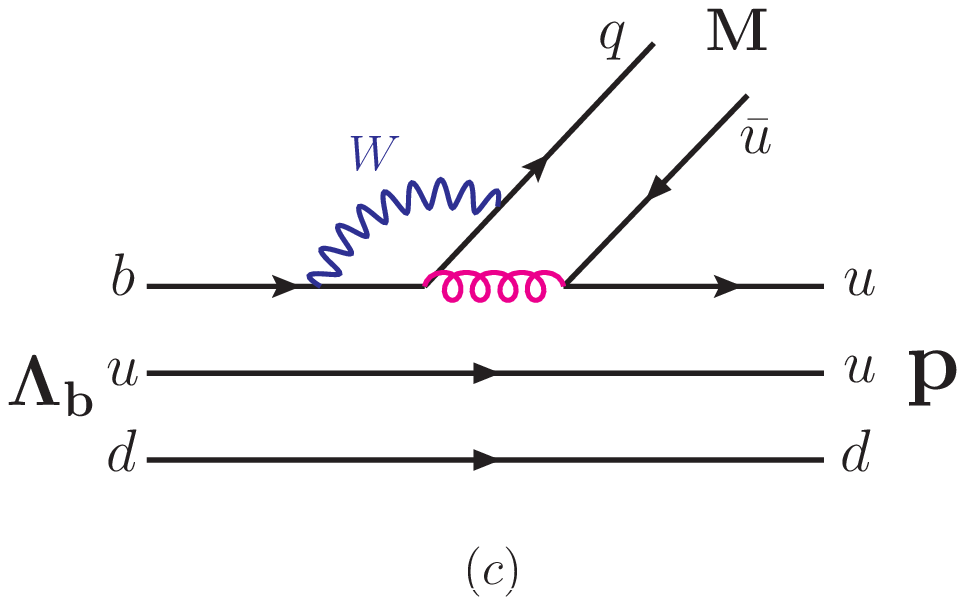}
\caption{Feynman diagrams depicted for 
(a) semileptonic $\Lambda_b\to (p,\Lambda_c)\ell\bar \nu_\ell$, 
(b) tree-level $\Lambda_b\to pM,\,\Lambda_c M_{(c)}$, and
(c) penguin-level $\Lambda_b\to p M$ decays.}\label{dia}
\end{figure}

As seen in Fig.~\ref{dia}, in terms of the quark-level effective Hamiltonian for 
the semileptonic $b\to u\ell\bar \nu_\ell$ and non-leptonic $b\to u\bar uq$ transitions, 
the amplitudes of the $\Lambda_b\to p\ell\bar \nu_\ell$ 
and $\Lambda_b\to  p M$ decays 
are found to be~\cite{Hsiao:2014mua,Detmold:2015aaa}
\begin{eqnarray}\label{amp1}
{\cal A}(\Lambda_b\to p\ell\bar \nu_\ell)&=&\frac{G_F}{\sqrt 2}V_{ub}
\langle p|\bar u\gamma_\mu(1-\gamma_5)b|\Lambda_b\rangle
\bar \ell\gamma^\mu(1-\gamma_5)\nu_\ell\,,\nonumber\\
{\cal A}(\Lambda_b\to p M)&=&i\frac{G_F}{\sqrt 2} f_M q^\mu\bigg[
\alpha_{M}\langle p|\bar u\gamma_\mu b|\Lambda_b\rangle+
\beta_{M}\langle p|\bar u\gamma_\mu\gamma_5 b|\Lambda_b\rangle\bigg]\,, 
\end{eqnarray}
where $G_F$ is the Fermi constant, $V_{ij}$ are the CKM quark mixing matrix elements, and
the matrix element of
$\langle M|\bar q\gamma^\mu(1-\gamma_5) u|0\rangle=if_M q^\mu$
corresponds to the meson production
with $f_M$ being the decay constant of $M$.
%
%
In Eq.~(\ref{amp1}), the parameter  $\alpha_{M}$ ($\beta_M$) is given by
$\alpha_{M}(\beta_{M})=V_{ub}V_{uq}^*a_1-V_{tb}V_{tq}^*(a_4\pm r_M a_6)$
with $r_M\equiv {2 m_M^2}/[m_b (m_q+m_u)]$
based on the factorization approach, where
$a_{1,4,6}\equiv c^{eff}_{1,4,6}+c^{eff}_{2,3,5}/N_c^{eff}$ 
with the effective Wilson coefficients $c_i^{eff}$ 
and color number $N_c^{eff}$~\cite{ali}, 
while $q=(d,s)$ for $M=(\pi^-,K^-)$. 
Similarly, the amplitudes of the $\Lambda_b\to \Lambda_c\ell\bar \nu_\ell$ 
and $\Lambda_b\to\Lambda_c M_{(c)}$ decays are given by
\begin{eqnarray}
\label{amp2}
{\cal A}(\Lambda_b\to \Lambda_c\ell\bar \nu_\ell)&=&\frac{G_F}{\sqrt 2}V_{cb}
\langle \Lambda_c|\bar c\gamma_\mu(1-\gamma_5)b|\Lambda_b\rangle
\bar \ell\gamma^\mu(1-\gamma_5)\nu_\ell\,,\nonumber\\
{\cal A}(\Lambda_b\to \Lambda_c M_{(c)})&=&
i\frac{G_F}{\sqrt 2}V_{cb}V_{\alpha\beta}^* 
a_1^{M_{(c)}}f_{M_{(c)}} q^\mu
\langle \Lambda_c|\bar c\gamma_\mu(1-\gamma_5)b|\Lambda_b\rangle\,,
\end{eqnarray}
where $V_{\alpha\beta}=V_{u(c)q}$ 
($q=d,s$) for $M_{(c)}=\pi^-(D^-),K^-(D^-_s)$ and 
$a_1^{M_{(c)}}$ are similar to $a_1$  in Eq.~(\ref{amp1}) but for the $M_{(c)}$ modes.
The amplitude of the $\Lambda_c^+\to \Lambda\ell\bar \nu_\ell$ decay 
through $c\to s\ell\bar \nu_\ell$ can be given 
by replacing $(b,c)$ with $(c,s)$ 
in ${\cal A}(\Lambda_b\to \Lambda_c^+\ell\bar \nu_\ell)$ of Eq.~(\ref{amp2}).
Note that the extractions with the $\Lambda_b\to pM$ and $\Lambda_b\to\Lambda_c M_{(c)}$ decays
are based on 
the validity of the factorization approach, 
which is supported by
the recent observations.
For example, with the factorization the ratios of 
${\cal B}(\Lambda_b\to \Lambda_c^+\pi^-)$/${\cal B}(\Lambda_b\to \Lambda_c^+ K^-)$,
${\cal B}(\Lambda_b\to \Lambda_c^+ D_s^-)$/${\cal B}(\Lambda_b\to \Lambda_c^+ D^-)$ and
${\cal B}(\Lambda_b\to p \pi^-)$/${\cal B}(\Lambda_b\to p K^-)$
are calculated to be 13.2, 25.1, and 0.84~\cite{Hsiao:2014mua,Hsiao:2015cda}, in agreement with
the data of $13.6\pm 1.6$, $24.0\pm 3.8$, and $0.84\pm 0.09$~\cite{pdg,r6}, respectively.
The other justification is from 
the soft-collinear effective theory~\cite{r8}. 
It is proposed that, if the factorization works,
the non-leptonic $\Lambda_b\to \Lambda_c^+\pi^-$ decay
can be related to the semileptonic $\Lambda_b\to\Lambda_c^+\ell\bar \nu_\ell$ decays,
resulting in the predictions of 
${\cal B}(\Lambda_b\to\Lambda_c^+\ell^-\bar \nu_\ell)\approx 6\times 10^{-2}$
and
${\cal B}(\Lambda_b\to\Lambda_c\pi^-)=4.6\times 10^{-3}$,
which remarkably agree with the data of  
$(6.2^{+1.4}_{-1.3})\times 10^{-2}$ and  
$(4.9\pm 0.4)\times 10^{-3}$ in the PDG~\cite{pdg}, respectively.

The amplitudes in Eqs.~(\ref{amp1}) and (\ref{amp2}) are related to
the matrix elements of the ${\bf B}_1\to {\bf B}_2$ transitions with
$({\bf B}_1,{\bf B}_2)=(\Lambda_b,p)$, $(\Lambda_b,\Lambda_c^+)$, 
and $(\Lambda_c^+,\Lambda)$
for the transition currents of $b\to u$, $b\to c$, and $c\to s$, respectively,
given by~\cite{Feldmann:2011xf}
\begin{eqnarray}\label{LbtoLc_ff}
&&\langle {\bf B}_2|\bar q_2 \gamma_\mu q_1|{\bf B}_1\rangle=
\bar u(p^\prime,s^\prime)\bigg[
f_0(q^2)(m_1-m_2)\frac{q^\mu}{q^2}
+f_+(q^2)\frac{m_1+m_2}{s_+}\nonumber\\
&&\times \bigg(p^\mu+p^{\prime\mu}-(m_1^2-m_2^2)\frac{q^\mu}{q^2}\bigg)
+f_\perp(q^2)\bigg(\gamma^\mu
-\frac{2m_2}{s_+}p^\mu-\frac{2m_1}{s_+}p^{\prime \mu}\bigg)
\bigg]u(p,s)\,,\nonumber\\
&&\langle {\bf B}_2|\bar q_2 \gamma_\mu\gamma_5 q_1|{\bf B}_1\rangle=
-\bar u(p^\prime,s^\prime)\gamma_5\bigg[
g_0(q^2)(m_1+m_2)\frac{q^\mu}{q^2}
+g_+(q^2)\frac{m_1-m_2}{s_-}\nonumber\\
&&\times \bigg(p^\mu+p^{\prime\mu}-(m_1^2-m_2^2)\frac{q^\mu}{q^2}\bigg)
+g_\perp(q^2)\bigg(\gamma^\mu
+\frac{2m_2}{s_-}p^\mu-\frac{2m_1}{s_-}p^{\prime \mu}\bigg)
\bigg]u(p,s)\,,
\end{eqnarray}
where $q=p-p'$, $s_\pm=(m_1\pm m_2)^2-q^2$,
and $f=f_j$ and $g_j$ ($j=0,+,\perp$) 
 are the form factors
in the helicity-based definition.
The momentum dependences of  
$f$
are written as~\cite{Detmold:2015aaa} 
\begin{eqnarray}
\label{LQCD_ff}
f(t)=\frac{1}{1-t/(m^f_{pole})^2}
\bigg[a_0^f+a_1^f\frac{\sqrt{t_+ -t_0}-\sqrt{t_+ -t_0}}{\sqrt{t_+ -t_0}+\sqrt{t_+ -t_0}}\bigg]\,,
\end{eqnarray}
where $m^f_{pole}$ are the pole masses and $t_0=(m_1-m_2)^2$, 
while $t_+$ and $a_{0,1}^f$ have been given in Refs.~\cite{Detmold:2015aaa,Meinel:2016dqj} .
Consequently,
one is able to integrate over the variables of the phase spaces
in the two and three-body decays for the decay widths~\cite{pdg}.

To demonstrate that 
the inclusions of ${\cal B}(\Lambda_b\to pM,\Lambda_c M_{(c)})$
in the extractions of $|V_{ub}|$ and $|V_{cb}|$
can reduce the theoretical uncertainties
from the $\Lambda_b\to (p,\Lambda_c)$ transition form factors 
due to the LQCD calculations in Eq.~(\ref{LbtoLc_ff}), 
we define
\begin{eqnarray}\label{F_cb}
{\cal F}(\Lambda_b\to{\bf B}_Q)_{q^2}=\frac{1}{|V_{Qb}|^2}
\int_{q^2}\,
\frac{\hat\tau_{\Lambda_b}}{(2\pi)^3\,32\,m_{\Lambda_b}^3}
\frac{d\Gamma(\Lambda_b\to{\bf B}_Q\ell\bar \nu_\ell)}{dq^2} dq^2\,,
\end{eqnarray}
which leads to that
${\cal B}(\Lambda_b\to {\bf B}_Q\ell\bar \nu_\ell)
=|V_{Qb}|^2 {\cal F}(\Lambda_b\to {\bf B}_Q)_{q^2>0\,\text{GeV}^2}$
with $q^2>(m_\ell+m_{\bar \nu})^2\simeq 0$~GeV$^2$,
where $Q=(c,u)$ for ${\bf B}_Q=(\Lambda_c,p)$ and
$\hat \tau_{\Lambda_b}\equiv \tau_{\Lambda_b}/(6.582\times 10^{-25})$.
Clearly, $R_{FF}$ defined in Ref.~\cite{Aaij:2015bfa} 
for the extraction of $|V_{ub}|/|V_{cb}|$
is in fact  the ratio of 
${\cal F}(\Lambda_b\to\Lambda_c)_{q^2>7\,\text{GeV}^2}$ to 
${\cal F}(\Lambda_b\to p)_{q^2>15\,\text{GeV}^2}$.

\section{Numerical Results and Discussions }

For the numerical analysis, we perform the minimum $\chi^2$ fit
with the experimental inputs given in Table~\ref{tab1},
where $|V_{ub}|$ and $|V_{cb}|$ are treated as the free parameters to be determined.
The theoretical inputs for the CKM matrix elements and decay constants
are given by~\cite{pdg}
\begin{eqnarray}\label{th_input}
(|V_{tb}|,10^3|V_{td}|,10^3|V_{ts}|)&=&(1.009\pm 0.031,8.2\pm 0.6,40.0\pm2.7)\,,\nonumber\\
(|V_{cd}|,|V_{cs}|)&=&(0.220\pm 0.005,0.995\pm 0.016)\,,\nonumber\\
(|V_{ud}|,|V_{us}|)&=&(0.97417\pm 0.00021,0.2248\pm 0.0006)\,,\nonumber\\
(f_\pi,f_K)&=&(130.2\pm 1.7,155.6\pm 0.4)\,\text{MeV}\,,\nonumber\\
(f_D,f_{D_s})&=&(203.7\pm 4.7,257.8\pm 4.1)\,\text{MeV}\,.
\end{eqnarray}
In addition, we use 
\begin{eqnarray}\label{th_input1}
a_1^{M_{(c)}}&=&1.1\pm 0.1\,,
\end{eqnarray}
which depends on the effective Wilson coefficients $(c_1^{eff},c_2^{eff})=(1.168,-0.365)$ 
and the effective color number $N_c^{eff}$.
 In Eq.~(\ref{th_input1}),  $N_c^{eff}$ has been taken to be $3$ as the central value,   
and ranging from 2 to $\infty$~\cite{ali} for the error
to account for  the non-factorizable effects in the generalized factorization.
Since $\Lambda_b\to p M$ have been tested to be
insensitive to the non-factorizable effects~\cite{Hsiao:2014mua}, 
we adopt the values of $a_{1,4,6}$ from Ref.~\cite{ali} with $N_c^{eff}=3$.
Note that the initial inputs for 
the $\Lambda_b\to (p,\Lambda_c)$ and $\Lambda_c\to \Lambda$ form factors 
defined in Eq.~(\ref{LQCD_ff}) are chosen  from Refs.~\cite{Detmold:2015aaa,Meinel:2016dqj}.

\begin{table}[b]
\caption{Inputs of the experimental data.
} \label{tab1}
\begin{tabular}{|c|c|c|}
\hline
&branching ratios&experimental data\\\hline
&
${\cal R}_{ub/cb}$
&$(0.95\pm 0.08)\times 10^{-2}$~\text{\cite{Amhis:2016xyh}}\\
$I1$ &$10^2 {\cal B}(\Lambda_b\to \Lambda_c\ell\bar \nu_\ell)$
&$6.2^{+1.4}_{-1.3}$~\text{\cite{pdg}}\\
&$\frac
{{\cal B}(\Lambda_b\to \Lambda_c\ell\bar \nu_\ell)}
{{\cal B}(\Lambda_c\to \Lambda \ell\bar \nu_\ell)}
$&$1.7\pm 0.4$~\text{\cite{pdg}}\\
\hline\hline
&$10^{6}{\cal B}(\Lambda_b\to p \pi^-)$
&$4.1\pm 0.8$~\text{\cite{pdg}}\\
$I2$ &$10^3 {\cal B}(\Lambda_b\to \Lambda_c\pi^-)$
&$4.6\pm 0.4$~\text{\cite{Aaij:2014jyk,pdg}}\\
&$10^4 {\cal B}(\Lambda_b\to \Lambda_c D^-)$
&$4.6\pm 0.6$~\text{\cite{pdg}}\\
\hline\hline
&$10^{6}{\cal B}(\Lambda_b\to p K^-)$
&$4.9\pm 0.9$~\text{\cite{pdg}}\\
$I3$ &$10^4 {\cal B}(\Lambda_b\to \Lambda_c K^-)$
&$3.6\pm 0.3$~\text{\cite{pdg}}\\
&$10^2 {\cal B}(\Lambda_b\to \Lambda_c D_s^-)$
&$1.1\pm 0.1$~\text{\cite{pdg}}\\
\hline
\end{tabular}
\end{table}
There can be two issues for the simultaneous extractions
of $|V_{ub}|$ and $|V_{cb}|$ in the exclusive $\Lambda_b$ decays.
First, when the non-leptonic and semileptonic decays are all included 
in the global fit, there are some possible uncertainties from the data correlations, 
which should be avoided or estimated. 
According to the {\sl ``CONSTRAINED FIT INFORMATION''} in the PDG~\cite{pdg},
${\cal B}(\Lambda_b\to\Lambda_c^+ \pi^-,\Lambda_c^+ K^-)$ and
${\cal B}(\Lambda_b\to p \pi^-,p K^-)$ 
are 94\% and 83\% correlated, respectively. Moreover,
${\cal B}(\Lambda_b\to\Lambda_c^+ \ell\bar \nu_\ell)$ 
has 14\% correlations with the individual value of 
${\cal B}(\Lambda_b\to\Lambda_c^+\pi^-)$ and ${\cal B}(\Lambda_b\to\Lambda_c^+ K^-)$.
As a result, we adopt 
${\cal B}(\Lambda_b\to\Lambda_c^+ \pi^-)=
(4.57^{+0.31}_{-0.30}\pm 0.23)\times 10^{-3}$
observed in Ref.~\cite{Aaij:2014jyk} and rescaled in the PDG~\cite{pdg}, 
instead of the weighted average one with other data,
to minimize  its correlation with 
${\cal B}(\Lambda_b\to\Lambda_c^+ \ell\bar \nu_\ell)$.
We also use three different scenarios with or without including
${\cal B}(\Lambda_b\to\Lambda_c^+ K^-,p K^-)$ in the fit to estimate the uncertainties.
Second, the recently updated data for ${\cal B}(\Lambda_c^+\to p K^-\pi^+)$
would help to data fitting as
$\Lambda_c^+$ is one of the final states.
In Table~\ref{tab1}, we have used 
the revised ${\cal R}_{ub/cb}$ value from Ref.~\cite{Amhis:2016xyh}, 
in which  the correction 
is around 5\%. 
Although ${\cal B}(\Lambda_b\to\Lambda_c^+ K^-)$ has an unknown
correction from ${\cal B}(\Lambda_c^+\to p K^-\pi^+)$~\cite{Aaij:2014lpa},
it has been excluded in one of the fitting scenarios.
On the other hand,
it is found that ${\cal B}(\Lambda_b\to \Lambda_c^+ \pi^-,\Lambda_c^+ D_s^-)$
and ${\cal B}(\Lambda_b\to \Lambda_c^+\ell\bar \nu_\ell,\Lambda_c^+ D^-)$
are 
free from ${\cal B}(\Lambda_c^+\to p K^-\pi^+)$,
whose examinations rely on the measurements in Refs.~\cite{Aaij:2014jyk,r5,r6}.
Accordingly,
the nine data points in Table~\ref{tab1} are classified 
into  three types of inputs, being denoted as $I1$, $I2$, and $I3$,
where $I1$ is for the semileptonic decays,
while $I2$ and $I3$ for the non-leptonic ones with $q=(d,s)$.
There can be three scenarios for the extractions.
In the first scenario ($S1$), we perform
the global fit with the six data points in $I1$ and $I2$ of Table~\ref{tab1},
such that there is no correlation in the calculation, leading to
\begin{eqnarray}\label{fit_VubVcb}
|V_{ub}|&=&(3.7\pm 0.3)\times 10^{-3}\,,\nonumber\\
(a_1^M,a_1^{M_c})&=&(1.14\pm 0.07,0.98\pm 0.06)\,,\nonumber\\
\chi^2/d.o.f&=&2.7/4\simeq 0.7\,,
\end{eqnarray}
where $d.o.f$ denotes as the degrees of freedom.
Note that $\chi^2/d.o.f\simeq 0.7$ 
in Eq.~(\ref{fit_VubVcb}) presents a very good fit. In addition,
$a_1^M$ and $a_1^{M_c}$ fitted with the slight deviations from the central value
and smaller errors than the initial one
imply the well-controlled non-factorizable effects.
Our fitting results for S1 are summarized in Table~\ref{tab2}, where we have also shown those
from LHCb and LQCD.
%
\begin{table}[b]
\caption{Fitting  results for the three scenarios of $S1$, $S2$, and $S3$, 
in comparison with the LHCb and LQCD ones.} \label{tab2}
{\footnotesize
\begin{tabular}{|c||c|c|c||c|c|}
\hline
&S1&S2&S3&LHCb~\text{\cite{Aaij:2015bfa}} &LQCD~\text{\cite {Detmold:2015aaa}}\\\hline
$10^3|V_{ub}|$
&$3.7\pm 0.3$
&$3.6\pm 0.2$
&$3.7\pm 0.4$
&$3.27\pm 0.15\pm 0.16\pm 0.06$
&-----\\
$10^3|V_{cb}|$
&$45.9\pm 2.7$
&$44.8\pm 2.0$
&$45.6\pm 3.7$
&-----
&-----\\
$\frac{|V_{ub}|}{|V_{cb}|}$
&$0.081\pm 0.008$
&$0.080\pm 0.006$
&$0.081\pm 0.011$
&$0.083\pm 0.004\pm 0.004$
&-----\\
${\cal F}(\Lambda_b\to\Lambda_c)_{q^2>0~{\text{GeV}^2}}$
&$31.16\pm 0.62$
&$31.25\pm 0.63$
&$31.22\pm 0.63$
&-----
&$31.19\pm 1.33$\\
${\cal F}(\Lambda_b\to\Lambda_c)_{q^2>7~{\text{GeV}^2}}$
&$12.17\pm 0.06$
&$12.18\pm 0.06$
&$12.18\pm 0.06$
&-----
&$12.17\pm 0.27$\\
${\cal F}(\Lambda_b\to p)_{q^2>0~{\text{GeV}^2}}$
&$37.99\pm 4.36$
&$37.90\pm 4.10$
&$37.37\pm 4.44$
&-----
&$37.41\pm 6.89$\\
${\cal F}(\Lambda_b\to p)_{q^2>15~{\text{GeV}^2}}$
&$18.11\pm 0.82$
&$18.08\pm 0.76$
&$17.90\pm 0.84$
&-----
&$17.92\pm 1.85$\\
$R_{FF}\equiv 
\frac{{\cal F}(\Lambda_b\to\Lambda_c)_{q^2>7~{\text{GeV}^2}}}
{{\cal F}(\Lambda_b\to p)_{q^2>15~{\text{GeV}^2}}}$
&$0.67\pm 0.03$
&$0.67\pm 0.03$
&$0.68\pm 0.03$
&$0.68\pm 0.07$
&$0.68\pm 0.07$\\
$10^4{\cal B}(\Lambda_b\to p \mu\bar \nu_\mu)$
&$5.2\pm 1.1$
&$4.9\pm 0.8$
&$5.1\pm 1.3$
&$4.1\pm 1.0$
&-----\\
\hline
\end{tabular}}
\end{table}

As shown in Table~\ref{tab2},
we obtain $|V_{cb}|=(45.9\pm 2.7)\times 10^{-3}$ in $S1$, which
agrees with the value of $(42.2\pm 0.8)\times 10^{-3}$ in Eq.~(\ref{data1})
from the inclusive $B$ decays. 
Furthermore, our result of 
$|V_{ub}|=(3.7\pm 0.3)\times 10^{-3}$ in Eq.~(\ref{fit_VubVcb})
is nearly identical to  that of $(3.72\pm 0.19)\times 10^{-3}$ in Eq.~(\ref{data1})
from the exclusive $B$ decays, but higher than 
$(3.27\pm 0.15\pm 0.16\pm 0.06)\times 10^{-3}$ from LHCb~\cite{Aaij:2015bfa}.
Compared to $|V_{cb}|=(39.5\pm 0.8)\times 10^{-3}$
from the exclusive $B$ decays, adopted by LHCb,
our extracted value of $|V_{cb}|$ has a lager uncertainty. Nonetheless,
we still get $|V_{ub}|$ with the error compatible to that of LHCb.
This is due to the fact that the measured ${\cal B}(\Lambda_b\to pM,\Lambda_c M_{(c)})$
are involved in the fitting, which reduce the theoretical uncertainties 
from $\Lambda_b\to (p,\Lambda_c)$ transition form factors
to be 2 times smaller than the value of
$0.68\pm 0.07$ in Refs.~\cite{Detmold:2015aaa,Aaij:2015bfa}.
It can be demonstrated by $R_{FF}=(0.67\pm 0.03,0.68\pm 0.07)$ 
from the fitting and the initial LQCD inputs,
respectively, where the nearly identical values of $R_{FF}$ 
show that LQCD calculation is also suitable for
the two-body $\Lambda_b$ decays that proceed at the low $q^2$ regions,
which have never been tested previously.
It is interesting to note that 
the connection of the fitted values of $|V_{ub}|$ and $|V_{cb}|$
causes $|V_{ub}|/|V_{cb}|=0.081\pm 0.008$, being nearly the same as
the value from the LHCb extraction.
We also predict 
${\cal B}(\Lambda_b\to p \mu\bar \nu_\mu)=(5.2\pm 1.1)\times 10^{-4}$,
which is slightly lager than $(4.1\pm 1.0)\times 10^{-4}$ by the extrapolation from
the data at $q^2>15$ GeV$^2$.

In the second scenario ($S2$),
we fit with all data points from $I1$, $I2$ and $I3$ in Table~\ref{tab1} with 
the correlations. Note that 
${\cal B}(\Lambda_b\to\Lambda_c K^-)$ also mixes with
the unknown contribution from ${\cal B}(\Lambda_c^+\to p K^-\pi^+)$.
As a result, we are  able to test possible deviations caused 
by the correlations as well as the new result from $\Lambda_c^+\to p K^-\pi^+$.
Furthermore, we  only use the three data points from the semileptonic processes 
in $I1$ 
as the third scenario ($S3$).
In this scenario, there is no need to introduce the factorization.
However, from Table~\ref{tab2} we find  that the fitted results in $S3$ 
are very close to those in $S1$, indicating that 
the correlations and the effect of $\Lambda_c^+\to p K^-\pi^+$ are insensitive to the fit.
We also see  that,
even without the factorization assumption,
the central value of $|V_{ub}|$ in $S3$ is almost the same as
those in $S1$ and $S2$, except  the larger errors. 
This implies that the global fit 
with the additional non-leptonic decays reduces
the uncertainties but without violating the outcome of the factorization.

\section{Conclusions}
We have performed the first simultaneous $|V_{ub}|$ and $|V_{cb}|$ extractions
in the exclusive $\Lambda_b$ decays. In addition to
the ratio of
${\cal B}(\Lambda_b\to p \mu\bar \nu_\mu)_{q^2>15\,\text{GeV}^2}$ to
${\cal B}(\Lambda_b\to \Lambda_c^+\mu\bar \nu_\mu)_{q^2>7\,\text{GeV}^2}$
measured by  LHCb,
the branching fractions of $\Lambda_b\to (p,\Lambda_c)\mu\bar \nu_\mu$,
$\Lambda_b\to pM$ and $\Lambda_b\to \Lambda_c M_{(c)}$
with $M_{(c)}=\pi^-(D^-)$ have been included in the global fit,
which help to eliminate the theoretical uncertainties 
from the $\Lambda_b\to (p,\Lambda_c)$ transition form factors, calculated in the LQCD model.
We have obtained $|V_{ub}|=(3.7\pm 0.3)\times 10^{-3}$, which is
 larger than the LHCb value of $(3.27\pm 0.15\pm 0.16\pm 0.06)\times 10^{-3}$
extracted from the $\Lambda_b$ decays also, but almost identical to that
 of $(3.72\pm 0.19)\times 10^{-3}$ from the exclusive $B$ decays.
In addition, our extracted result of $|V_{cb}|=(45.9\pm 2.7)\times 10^{-3}$ is close to 
 $(42.2\pm 0.8)\times 10^{-3}$ from the inclusive $B$ decays.
We have predicted
${\cal B}(\Lambda_b\to p \mu\bar \nu_\mu)=(5.2\pm 1.1)\times 10^{-4}$,
in comparison with
the extrapolated value of $(4.1\pm 1.0)\times 10^{-4}$
from the partial branching ratio of $\Lambda_b\to p \mu\bar \nu_\mu$
at $q^2>15$ GeV$^2$.

\section*{ACKNOWLEDGMENTS}
This work was supported in part by National Center for Theoretical Sciences,
MoST (MoST-104-2112-M-007-003-MY3), and
National Science Foundation of China (11675030).

\end{document}